# Calculations of the Propagated LIS Electron Spectrum
# Which Describe the Cosmic Ray Electron Spectrum below ~100 MeV
# Measured Beyond 122 AU at Voyager 1 and its Relationship to the PAMELA
# Electron Spectrum above 200 MeV

W.R. Webber[1] and P.R. Higbie[2]

1. New Mexico State University, Astronomy Department, Las Cruces, NM 88003, USA
2. New Mexico State University, Physics Department, Las Cruces, NM  88003, USA

# ABSTRACT

The new Voyager measurements of cosmic ray electrons between 6-60 MeV beyond 122 AU are very sensitive indicators of cosmic ray propagation and acceleration in the galaxy at a very low modulation level. Using a Monte Carlo diffusion model with a source spectrum with a single spectral index of -2.2 at all energies we are able to fit this observed Voyager spectrum and the contemporary PAMELA electron spectrum over an energy range from 6 MeV to ~200 GeV. This spectrum has a break in it but this break is due to propagation effects, not changes in the primary spectrum. This break is gradual, starting at $\geq 2$ GeV where the spectrum is $\sim E^{-3.2}$ and continuing down to ~100 MeV or below where the spectrum becomes $\sim E^{-1.5}$. At the higher energies the loss terms due to synchrotron radiation and inverse Compton effects which are $\sim E^{2.0}$ steepen the exponent of the source spectrum by 1.0. At lower energies, these terms become unimportant and the loss is governed by diffusion and escape from the galaxy. A diffusion term which is proportional to $\beta^{-1}$ below ~0.32 GV (which also fit the H and He spectra measured at Voyager) and has a value = $3 \times 10^{28}$ cm$^2 \cdot$s$^{-1}$ at 1 GV and a boundary at $\pm 1$ Kpc will fit the Voyager or other similar spectra at low energies.

**Introduction**

At energies below a few GeV, down to a few MeV, the effects of solar modulation in the heliosphere greatly distort the LIS galactic electron spectrum and are difficult to correct for. These modulation effects have been found to be a factor ~100 times or more between a possible LIS electron spectrum and those measured near the Earth below 100 MeV as determined by Voyager measurements (McDonald, et al., 2012). In addition, the galactic propagation of electrons at these lower energies is no longer dominated by E loss processes related to the B field and other processes which are proportioned to E or $E^2$, but is more controlled by diffusion processes and their energy or rigidity dependence. As a result the LIS electron energy spectra now flatten from their spectra with an exponent between -3.1 and -3.3 that has been observed above a few GeV to an exponent of about -1.5, depending on the rigidity dependence of the diffusion coefficient and other factors (see e.g., Webber and Higbie, 2008; Strong, et al., 2011).

There has been too much uncertainty to be able to make a reliable calculation of the demodulated LIS electron spectrum from measurements at the Earth below 1-2 GeV. But the remarkable Voyager spacecraft has again come to the rescue of its Earthbound creators. After August 25, 2012, this spacecraft passed through a region of extent <1 AU at about 122 AU from the Sun during which the energetic MeV radiation in the heliosheath decreased by factors of up to 1000 and the intensities of galactic nuclei and electrons jumped up to levels up to a factor ~2 times those observed earlier (Webber and McDonald, 2013; Stone, et al., 2013). These levels corresponded to near zero solar modulation, from modulation levels corresponding to possibly 80 MV just 1 AU earlier (on this scale the modulation level at the Earth would be 270 MV) (see Webber, Higbie and McDonald, 2013a).

This new Voyager electron spectrum from Stone, et al., 2013, for this time period is shown in Figure 1 along with earlier calculations of possible IS electron spectra by Langner, et al., 2001 and Webber and Higbie, 2008. For our new calculation of electrons in this paper we are considering this Voyager spectrum to be a local IS spectrum. It is thus an extension to low energies of a demodulated PAMELA spectrum measured above a few 100 MeV also shown in Figure 1. The earlier calculations shown in Figure 1 do a good job of bracketing the low energy LIS electron spectrum within a factor $\pm$ 2 and show how the spectrum changes with energy.

Our first step in this paper will then be to compare this new Voyager electron spectrum from ~6-60 MeV with updated models for the propagation of electrons in the galaxy as carried out earlier by Webber and Higbie, 2008, but now extending down to a few MeV and to determine the energy (rigidity) dependence of these changing features. We will then compare the Voyager electron spectrum with that measured by PAMELA and discuss modulation effects.

**Propagation of Electrons in the Galaxy**

Here we employ a Monte Carlo Diffusion Model (MCDM) to study the propagation of electrons (and nuclei) in the Galaxy. This model is described more completely in Webber [1993] and Webber and Rockstroh [1997] and Webber and Higbie [2008]. The model is a one-dimensional transport equation for diffusion out of the disk of the galaxy. It has a diffusion coefficient $K(z)$ and a source function $S(z)$ that includes a cosmic ray source and secondary production (and loss) by nuclear interactions and other appropriate energy loss (or gain) processes which are dependent on the species:

$$\partial n/\partial t = S(z) + \partial/\partial z \, [K(z) \, \partial n/\partial z]$$

The diffusion parameters for electron propagation are initially derived from the application of this same program to the propagation of protons and heavier nuclei as, for example, the B/C ratio [Webber, 2000]. Specifically in the earlier calculation by Webber and Higbie, 2008, the diffusion coefficient was taken to be $3.0 \cdot 10^{28}$ $(P/P_0)^{0.5}$ cm$^2$ s$^{-1}$ above $P_0 = 3.0$ GV, and equal to a const $= 3.0 \cdot 10^{28}$ cm$^2$ s$^{-1}$ below 3.0 GV. A diffusive halo of thickness $= \pm Z_B = 1$ Kpc is assumed and a disk matter column density integral, $I_m = \int n dz = 8.5 \cdot 10^{20}$ cm$^2$, where $n_0 = 1.2$ cm$^3$ and the Z dependence of the total matter density in the disk is given by exp- $(Z/Z_m)$ where $Z_m = 0.2$ Kpc. At $\pm Z_B = 1$ Kpc the electrons are assumed to freely escape the extended disk. For electrons, specifically, the B field magnitude as a function of Z taken to b$_e$ exp- $(Z/Z_B)$ where $Z_B = \pm 1.5$ Kpc. The energy loss by synchrotron radiation $(dE/dt) = -b_s \, E^2$ where $b_{so} = 2 \cdot 10^{-16}$ s$^{-1}$ ($B_0 = 5$ µG). The inverse Compton energy loss is described by $dE/dt = -b_{ICo} \cdot E^2$ where the value of $b_{ICo} = 0.5 \cdot 10^{-16}$ s$^{-1}$ and the energy loss by bremsstrahlung and ionization are determined from the matter distribution described above. The cosmic ray source is assumed to be at $Z = 0$.

Note that below about 1 GeV the above loss terms for electrons which are proportional to E or $E^2$ become small and the propagation of electrons is determined mainly by the diffusion term and its energy dependence.

In Figure 2 we show the newly calculated LIS spectra (x $E^2$) for electrons obtained using source spectra with a spectral index = -2.2. Source spectra with indices less than -2.1 or greater than -2.3 are either too flat or too steep to fit the electron spectra measured by Voyager and PAMELA simultaneously. The measured electron spectrum from Voyager is shown in red in Figure 2 and is from Stone, et al., 2013. Also shown in Figure 2 is the electron spectrum near Earth obtained by PAMELA (Adriano, et al., 2011; Potgieter and Nndanganemi, 2013) in 2009. Our new LIS spectra should be compared with others, here we take as examples the spectra presented in Figure 1 by Langner, 2001, and Webber and Higbie, 2008.

For the new propagation calculations in Figure 2 the rigidity, $P_0$, at which the galactic diffusion coefficient changes from it high rigidity dependence, $P^{0.5}$, to the low rigidity dependence for the diffusion coefficient of $P^{-1.0}$ used in this paper is varied from 1.0 GV to zero. A value of $P_0 = 0.32$ gives a good fit to the new Voyager spectrum and also the PAMELA spectrum as can be seen in Figure 2. The reason for this ability to fit both spectra is that the E loss terms which are $\sim E^2$ steepen the source spectrum of index $\sim 2.2$ by a power $\sim 1.0$ above 1-2 GeV where they become important, thus producing the -3.1 - -3.2 spectrum that PAMELA observes at higher energies.

The numerical values of the rigidity dependence of the diffusion coefficient are shown in Figure 3, along with some earlier example of the dependence used by Webber and Higbie, 2008, including their calculations shown in Figure 1 which are the Monte Carlo Diffusion Model, MCDM (1) and (2). These same diffusion coefficients used here also apply to the H and He spectra that are calculated in Webber, Higbie and McDonald, 2013b, that fit the Voyager 1 measurements of these nuclei (Stone, et al., 2013).

**The LIS Electron Spectrum from 0.2-2.0 GV, Solar Modulation and the PAMELA Proton and Electron Spectra Measured in 2009**

For this modulation calculation we make use of the fact that both Voyager and PAMELA have measured the proton spectrum over a corresponding energy range. We also assume that the proton and electron spectra measured by Voyager in late 2012 are what is commonly referred to as the LIS. In this way we can obtain a modulation ratio, LIS/Earth, for protons which corresponds to the level of modulation at the Earth in 2009. This measured ratio may then be compared with a calculated ratio based on the LIS used in the calculation of the proton spectrum to be expected at the Earth by Webber, Higbie and McDonald, 2013a, in a two-zone modulation model where the overall modulation potential between LIS and Earth is taken to be 270 MV. We can then repeat this procedure for electrons and in so doing obtain a new better validated estimate of the LIS spectrum in the lower energy range covered by PAMELA, e.g., 0.25-2.0 GeV. A similar approach has been suggested by Menn, et al., 2013. In what follows we discuss in more detail our procedure.

On the PAMELA experiment, both the proton and electron spectra have been precisely measured over the rigidity range ~0.2 to 2.0 GV in 2009, at the time of the lowest solar modulation levels ever recorded at the earth, estimated to be between 250-270 MV (e.g., Wiedenbeck, 2012).

The proton spectrum at Voyager has been measured over approximately the same rigidity range as PAMELA. As a result we can utilize the comparison of intensities for protons measured at V1 and at PAMELA to obtain a ratio $j_p(V)/j_p(P)$. This may be considered as the total modulation between the Earth and the LIS. The $\ell$n of this measured ratio is shown in Figure 4 as red points. The calculated ratio for a solar modulation =270 MV is shown as a red line (see Webber, Higbie and McDonald, 2013a). The agreement between the model predictions and the actual measurements of this quantity is excellent.

The electron spectrum over roughly the same range of rigidities as the protons has also been well measured at the same time by PAMELA. The measurements of electrons at Voyager are at lower rigidities than PAMELA, but we may use the calculated LIS spectrum for $P_0 = 0.32$ GV in Figure 2 in the range 0.2-2 GV, which also fits the lower rigidity Voyager electrons and then take the ratio $j_p(V)/j_p(P)$. These ratios are shown as black points in Figure 4 at the rigidities at which the electron measurements are made by PAMELA.

The black curve in Figure 4 is the calculated ratio $j_e(V)/j_e(P)$ from the same modulation model and set of diffusion parameters that are used for protons. Here the argument is good, also. These total modulation curves for both protons and electrons provides a new level of confidence for the modeling of the total GCR modulation in the heliosphere.

## Summary and Conclusions

In August 2012, Voyager entered a new region of space in which the galactic electron intensity from 5-60 MeV being monitored by the TET electron detector (Stone, et al., 1977), suddenly increased a factor ~2 to record levels. This new electron intensity beyond ~122 AU is several hundred times the typically measured intensity at these same energies at the Earth and typical of values of calculated LIS intensities in several propagation models. Most of this increase in electrons has occurred in the heliosheath; however; we note that after entering this new region in August 2012 the electron intensity has remained constant to within $\pm$1-2% now for ~ 1 year (corresponding to a 3.6 AU outward movement of V1).

We have assumed that this electron intensity corresponds to the LIS intensity and, updating our earlier Monte Carlo diffusion models, have been able to reproduce the Voyager electron intensity and spectrum measured at this time. This is achieved using a galactic source spectrum with index ~-2.20, independent of energy, and a diffusion coefficient ~$P^{0.5}$ above a rigidity ~0.32 GV, below which the diffusion coefficient becomes ~$P^{-1.0}$. This calculated LIS electron spectrum also accurately ($\pm$ 10%) fits the higher energy PAMELA electron data obtained in 2009 from 0.2 to up to 200 GeV when the electron modulation is included using an approach based on Voyager and PAMELA proton data. This approach involves comparing the Voyager and PAMELA measurements of both protons and electrons assuming that the Voyager measurements are the LIS spectra. These values are then compared with modulation calculations. In so doing we have obtained a determination of the modulation fraction of protons between the LIS and the Earth in 2009 that will be useful for all atmospheric studies which rely on the GCR input spectrum. For electrons this same procedure is used to provide a LIS electron spectrum and a heliospheric modulation fraction in the energy range 0.25-2.0 GeV.

The ability to reproduce the observed Voyager and PAMELA electron spectra with a single power law rigidity spectrum with index ~-2.2 when propagation and modulation effects are taken into account, is significant. The fact that this power law is almost the same as the single index of -2.24 that fits all three of the spectra of H, He and C/O nuclei measured at V1 at the same time along with higher energy data for the same nuclei, from a few MeV to ~10 GeV, when the propagation effects are taken into account (see Webber, Higbie and McDonald, 2013b) makes a spectral index of between ~2.2-2.25 a robust indicator of the spectra of GCR in this energy range.

**Acknowledgments:** We acknowledge the programming work of Jason Peterson and Ricky Webber, without which this program could not have been improved and extended to lower energy. This effort is supported by JPL-Voyager to which we are grateful. Our P.I. is Ed Stone who has ably directed our team for almost 40 years. The electron spectrum is due to the work of Alan Cummings and Nand Lal. But none of this would have been possible without the constant prodding of Frank McDonald and the graphs that Bryant Heikkila and others prepared.

**FIGURE CAPTIONS**

**Figure 1:**  Voyager 1 electron spectra measured by Stone, et al., 2013 (in red), and also the 2009 PAMELA electron spectrum, along with earlier predictions of the LIS electron spectrum by Langner, de Jager and Potgieter, 2001, and Webber and Higbie, 2008,  as discussed in the text.

**Figure 2:**  Calculated LIS electron spectra (x $E^2$) in this paper using the MCDM with a source index =-2.20 and a diffusion coefficient =$P^{0.5}$ above $P_0$ and ~$P^{-1.0}$ below $P_0$.  $P_0$ varies from 0.176 to 1.0.  The Voyager 1 measurements and the PAMELA 2009 measurements are shown in red.

**Figure 3:**  Numerical values of the diffusion coefficients and their rigidity dependence used in this paper (in red).  Earlier values used by Webber and Higbie, 2008, shown in black.

**Figure 4:**  Ratio of the measured Voyager proton spectrum (Stone, et al., 2013), to the proton spectrum measured by PAMELA (Adriano, et al., 2013), at the maximum in 2009 (solid red circles).  The red line is the calculated ratio of the LIS spectrum to that at the Earth for protons for a total modulation potential of 270 MV.  The black line is the same ratio calculated for electrons for the same modulation potential of 270 MV.  The black open circles show this ratio for the calculated LIS electron intensities for $P_/$=0.32 GV in Figure 2 to the measured electron intensities at PAMELA in 2009.

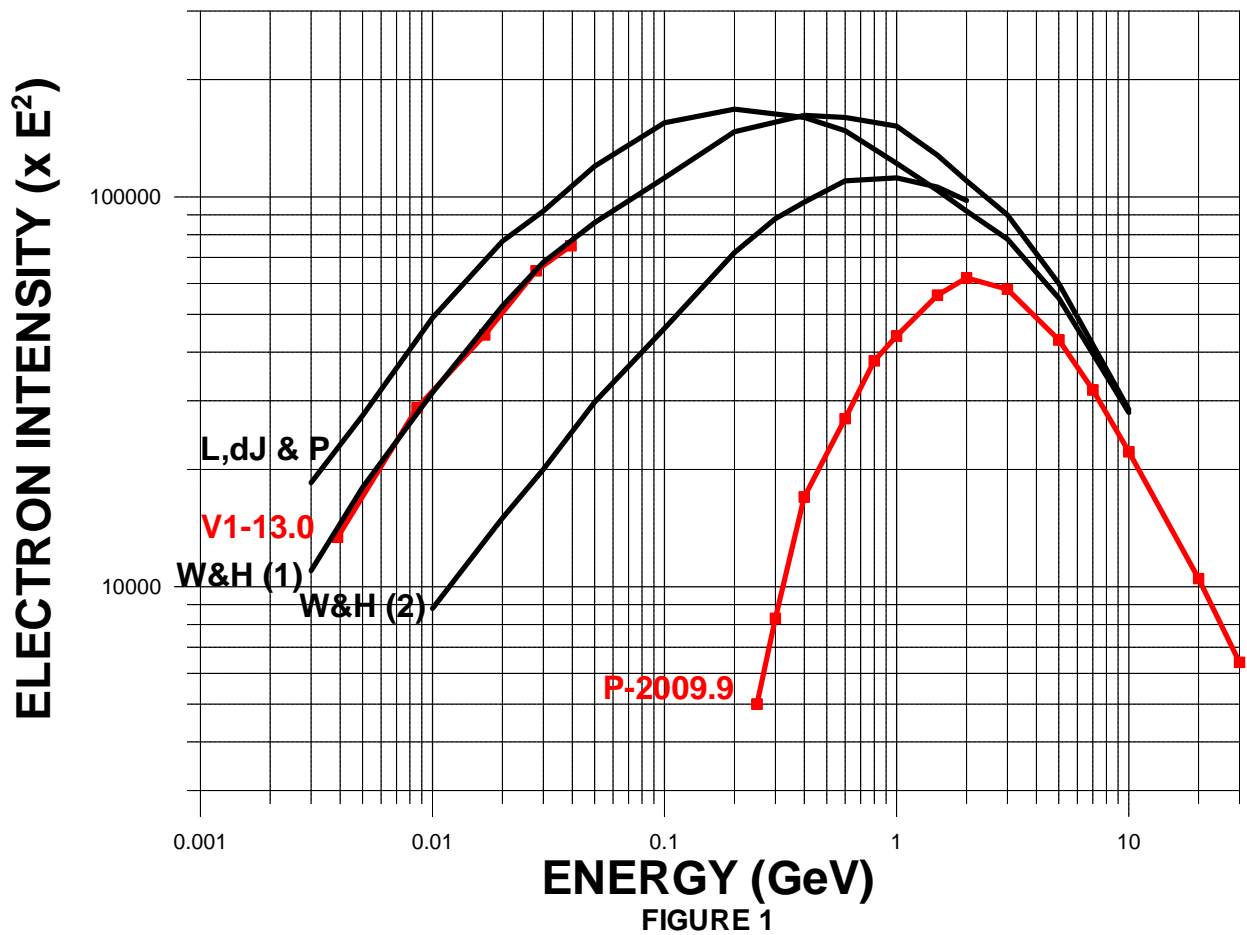

FIGURE 1

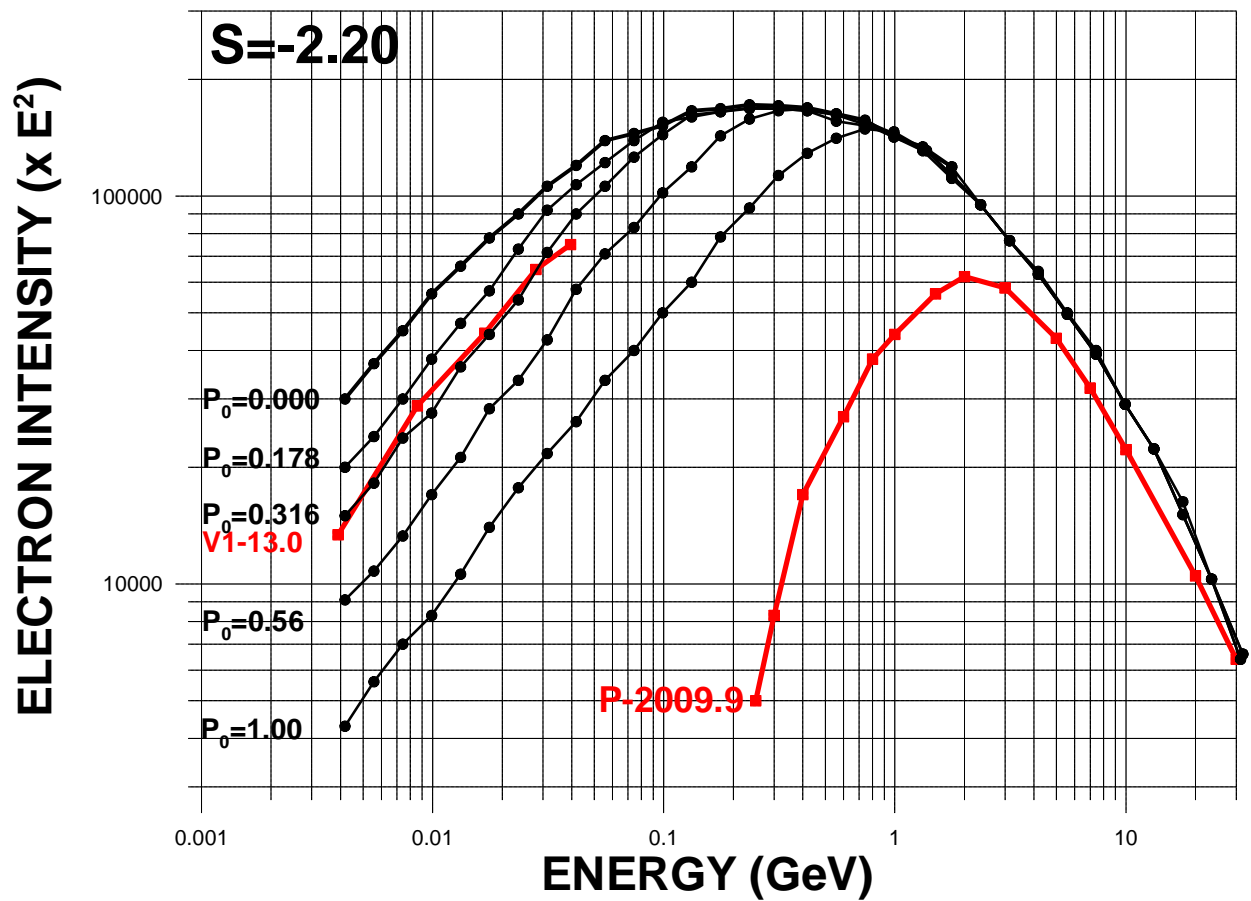

FIGURE 2

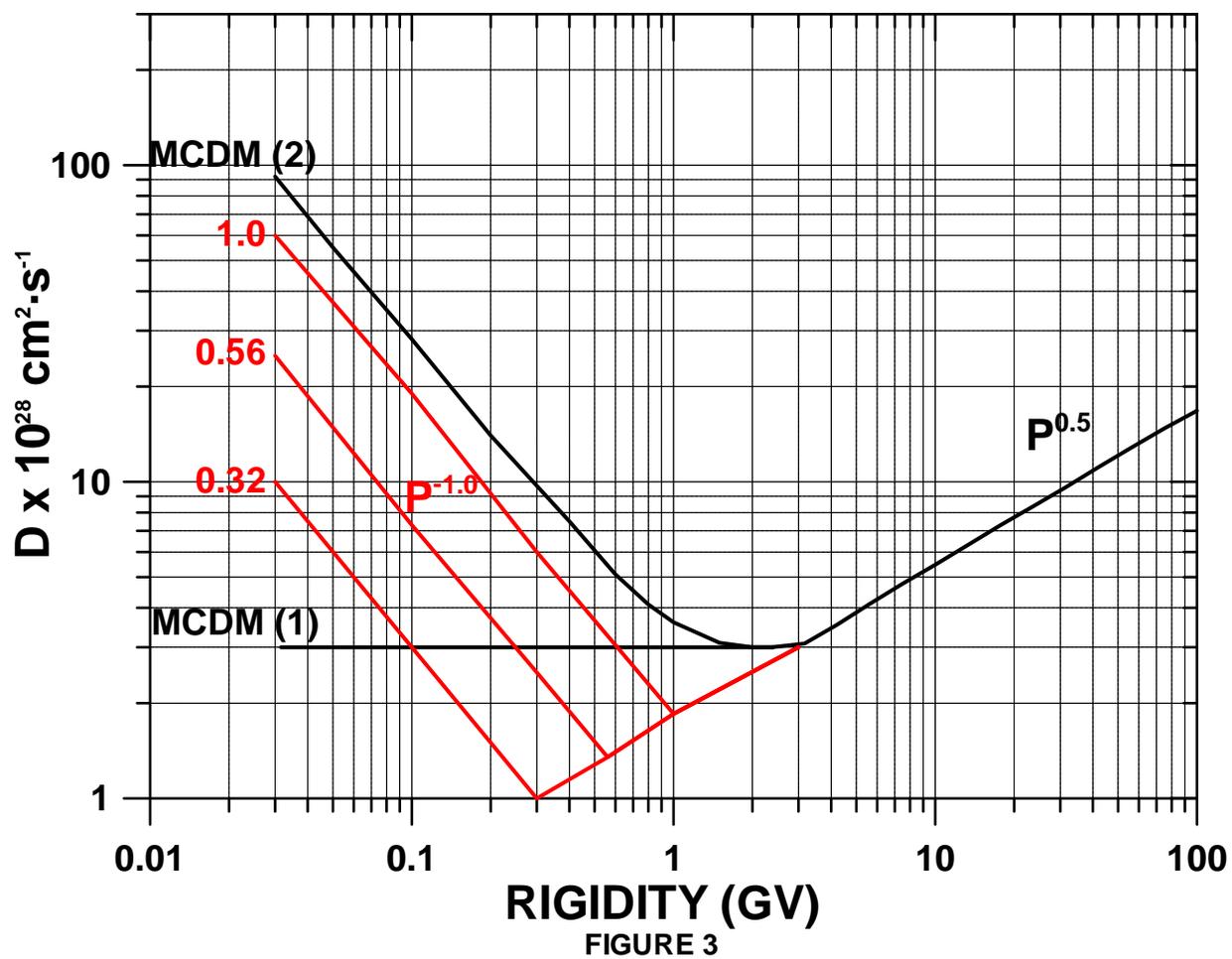

FIGURE 3

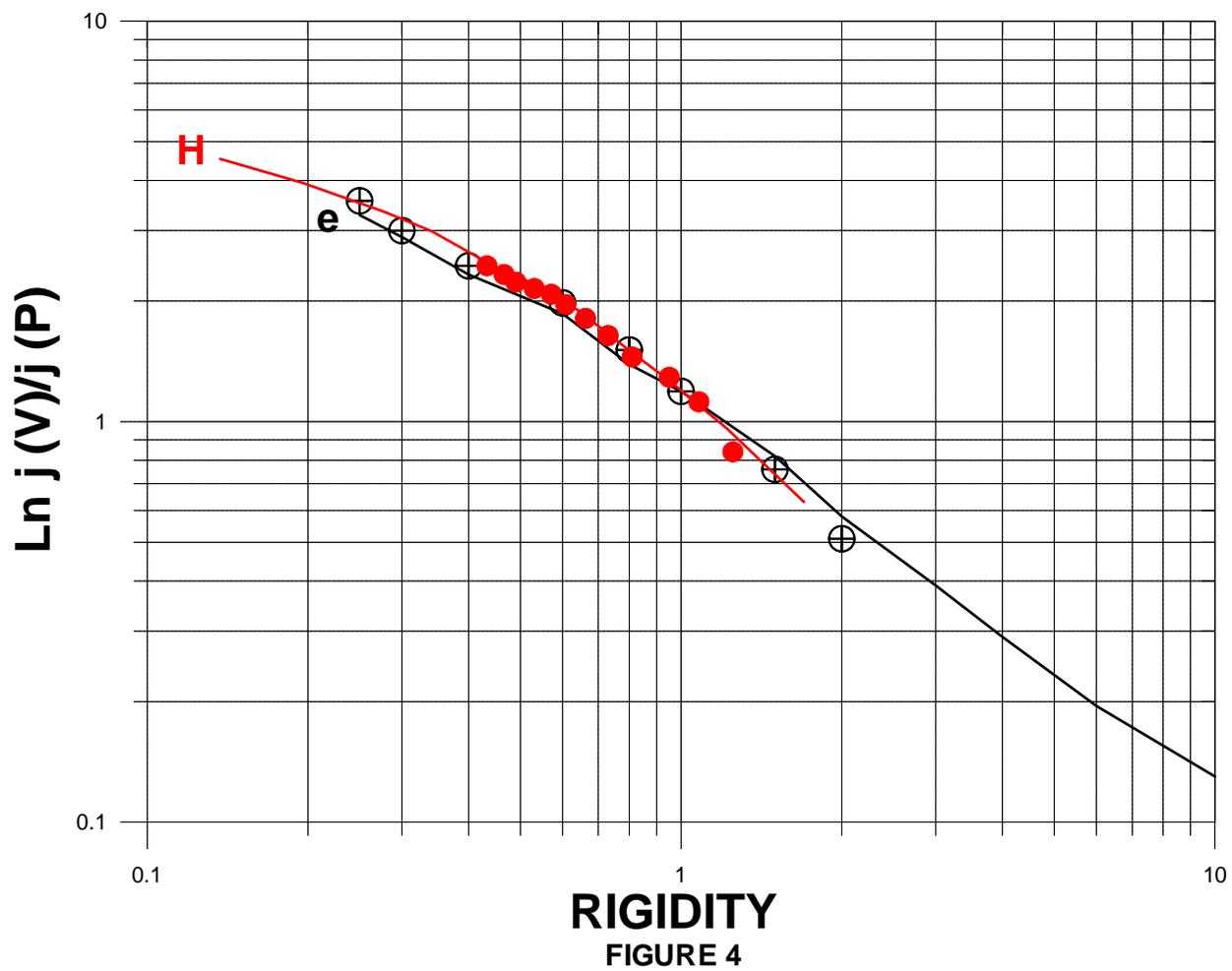

**FIGURE 4**